\documentclass[english]{article}
\usepackage[T1]{fontenc}
\usepackage[latin9]{inputenc}
\usepackage{setspace}
\usepackage{amssymb}

\makeatletter
\newcommand{\lyxaddress}[1]{
\par {\raggedright #1
\vspace{1.4em}
\noindent\par}
}

\makeatother

\usepackage{babel}

\begin{document}
\begin{onehalfspace}

\title{Dark matter from {}``strong gravity'' - consistent with CRESST,
CoGeNT and DAMA/LIBRA}
\end{onehalfspace}

\begin{onehalfspace}

\author{T. R. Mongan}
\end{onehalfspace}

\maketitle
\begin{onehalfspace}

\lyxaddress{84 Marin Avenue, Sausalito, CA 94965 USA}

\lyxaddress{tmongan@gmail.com}
\end{onehalfspace}
\begin{abstract}
\begin{onehalfspace}
Kelso, Hooper and Buckley {[}arXiv:1110.5338{]} found CRESST, CoGeNT
and DAMA/LIBRA results are consistent with 10 - 15 GeV dark matter
particles. Hennawi and Ostriker {[}arXiv:astro-ph/0108203{]} analyzed
supermassive black hole formation in the centers of galaxies, finding
a best fit for dark matter (self-interaction cross-section)/mass ratio
= 0.02 cm$^{2}$/g, with round-off error $\pm$25\%. Combining the
Hennawi/Ostriker result with the {}``strong gravity'' model for
dark matter {[}arXiv:0706.3050{]} requires dark matter particles with
mass between 10.5 GeV and 17.5 GeV, overlapping the Kelso/Hooper/Buckley
dark matter particle mass range.\end{onehalfspace}

\end{abstract}
\begin{onehalfspace}

\subsection*{Introduction}
\end{onehalfspace}

\begin{onehalfspace}
The Kelso, Hooper and Buckley \cite{key-1} analysis of CRESST, CoGeNT
and DAMA/LIBRA results indicates a dark matter particle mass of 10
- 15 GeV. Based on the Kelso/Hooper/Buckley work, the following analysis
shows CRESST, CoGeNT and DAMA/LIBRA results and the Hennawi/Ostriker
\cite{key-2} estimate of (self-interaction cross-section)/mass ratio
of dark matter particles, are consistent with dark matter from {}``strong
gravity'' \cite{key-3} 

Hennawi and Ostriker's \cite{key-2} best fit of $\frac{\sigma}{M}$
= 0.02 cm$^{2}$/g for the (self-interaction cross-section)/mass ratio
of dark matter particles is based on growth of black holes in the
center of galaxies. The Hennawi/Ostriker estimate for $\frac{\sigma}{M}$
is consistent with upper bounds on $\frac{\sigma}{M}$ based on evaporation
of galactic halos \cite{key-4}, optical and X-ray observations of
the colliding {}``Bullet Cluster'' galaxies 1E0657-56 \cite{key-5}
\cite{key-6}, and X-ray observations of the merging galaxy cluster
MACSJ0025.4-1222 \cite{key-7}.
\end{onehalfspace}

\begin{onehalfspace}

\subsection*{Dark Matter}
\end{onehalfspace}

\begin{onehalfspace}
Many cosmological models assume all four forces governing the universe
were unified very early in the history of the universe. After the
initial force symmetry broke, the gravitational structure constant
$\frac{Gm_{p}^{2}}{\hbar c}=5.9\times10^{-39}$, with $\hbar=1.05\times10^{-27}$
g cm$^{2}$/sec, $c=3\times10^{10}$cm/sec and $m_{p}=1.67\times10^{-24}g$,
is the ratio of the strength of gravity and the strong force after
inflation. In the flat homogeneous and isotropic space of the post-inflationary
universe with matter energy density $\rho$, the {}``strong gravity''
model for dark matter \cite{key-3} approximates the strong force
as an effective \textquotedbl{}strong gravity\textquotedbl{} (acting
only on matter) with strength $G_{S}=\left(\frac{M_{P}}{m_{p}}\right)^{2}G=1.7\times10^{38}G$,
where the gravitational constant $G=6.67\times10^{-8}$cm$^{3}$/g
sec$^{2}$ and the Planck mass $M_{P}=\sqrt{\frac{\hbar c}{G}}=2.18\times10^{-5}$
g. Then, the strong gravity Friedmann equation $\left(\frac{dR}{dt}\right)^{2}-\frac{8\pi G_{S}\rho R^{2}}{3}=-c^{2}$
describes the local curvature of spaces defining closed massive systems
bound by the effective strong gravity. Because a strong force at short
distances is involved, a quantum mechanical analysis of such systems
is necessary. The Schrodinger equation resulting from Elbaz-Novello
quantization of the Friedmann equation \cite{key-8} for a closed
massive system bound by the effective strong gravity is $-\frac{\hbar^{2}}{2\mu}\frac{d^{2}}{dr^{2}}\psi-\frac{2G_{S}\mu M}{3\pi r}\psi=-\frac{\mu c^{2}}{2}\psi$,
where $M=2\pi^{2}\rho r^{3}$ is the conserved mass of the closed
system with radius $r$ and $\mu$ is an effective mass. This Schrodinger
equation is identical in mathematical form to the Schrodinger equation
for the hydrogen atom and can be solved immediately. The ground state
curvature energy $-\frac{\mu}{2\hbar^{2}}\left(\frac{2G_{S}\mu M}{3\pi}\right)^{2}$
of this Schrodinger equation must equal $-\frac{\mu c^{2}}{2}$ for
consistency with the corresponding Friedmann equation, so the effective
mass $\mu=\frac{3\pi\hbar c}{2G_{S}M}$. The ground state solution
of this Schrodinger equation describes a stable closed system bound
by the effective strong gravity, with zero orbital angular momentum
and radius $<r>=\frac{G_{S}M}{\pi c^{2}}=\frac{\hbar M}{\pi cm_{p}^{2}}$.
Geodesic paths inside the stable ground state closed systems created
by the effective strong gravity are all circles with radius $<r>=\frac{\hbar M}{\pi cm_{p}^{2}}$,
so matter within these closed systems is permanently confined within
a sphere of radius $<r>$. No particle can enter or leave these small
closed systems after they form, to increase or decrease the amount
of matter in those closed systems. These small closed systems act
like rigid impenetrable spheres interacting only gravitationally and
constituting the majority of dark matter.

Assuming velocity-independent rigid sphere scattering \cite{key-9},
the (self-interaction collision cross-section)/mass ratio for the
dark matter particles is $\frac{\sigma}{M}=\frac{4\pi\left(2r\right)^{2}}{M}$.
Consider values of $\frac{\sigma}{M}$ between 0.015 cm$^{2}$/g and
0.025 cm$^{2}$/g, that round off to the Hennawi/Ostriker estimate
of $\frac{\sigma}{M}$ = 0.02 cm$^{2}$/g. Inserting the dark matter
particle radius/mass relation $r=\frac{\hbar M}{\pi cm_{p}^{2}}$
into the rigid sphere (self-interaction collision cross-section)/mass
relation $\frac{\sigma}{M}=\frac{4\pi\left(2r\right)^{2}}{M}$ yields
$M=\left[\left(\frac{\sigma}{M}\right)\frac{\pi}{16}\left(\frac{c}{\hbar}\right)^{2}m_{p}^{3}\right]m_{p}.$
Therefore, values of $\frac{\sigma}{M}$ between 0.015 cm$^{2}$/g
and 0.025 cm$^{2}$/g indicate a dark matter particle mass between
10.5 GeV and 17.5 GeV, overlapping the 10 - 15 GeV dark matter particle
mass range determined by the Kelso/Hooper/Buckley \cite{key-1} analysis
of CRESST, CoGeNT and DAMA/LIBRA results. The nucleon mass equivalent
$A=\left(\frac{\sigma}{M}\right)\frac{\pi}{16}\left(\frac{c}{\hbar}\right)^{2}m_{p}^{3}$
of the dark matter particles ranges from 11.2 to 18.7, and the radius
of the dark matter particles $r=\left(\frac{A}{\pi}\right)\left(\frac{\hbar}{m_{p}c}\right)$
ranges from 0.75 fm to 1.25 fm.

The radius of the dark matter particles, estimated from the Hennawi/Ostriker
$\frac{\sigma}{M}$ result, relates to the finite range of the strong
force. The distance $\pi r$, halfway around a geodesic path inside
the small closed systems constituting most dark matter, is a characteristic
length for the collective strong gravity effective force binding $A$
nucleon mass equivalents into dark matter particles. It's roughly
analogous to the radius of an atomic nucleus containing $A$ nucleons,
bound by collective effects of the strong interaction. The characteristic
length for dark matter particles $\pi r=A\left(\frac{\hbar}{m_{p}c}\right)$,
ranging from 2.4 fm to 3.9 fm, is similar to the 2.4 fm to 2.9 fm
range of root mean square charge radii \cite{key-10} for nuclei containing
10 (boron) to 19 (fluorine) nucleons.

The radius of a nucleus comprised of $A$ nucleons is often approximated
as $r_{n}=1.25A^{\frac{1}{3}}$ fm. Considering only nuclei comprised
of 10 to 19 nucleons, a least squares fit to ten tabulated values
\cite{key-10} of root mean square charge radii gives $r_{n}=1.04A^{0.34}$
fm $\left[R^{2}=0.91,\: p=2\times10^{-5}\right]$. If these relations
adequately approximate the effective length $\pi r=A\left(\frac{\hbar}{m_{p}c}\right)$
characterizing $A$ nucleon mass equivalents bound into a dark matter
particle, they suggest a dark matter particle mass of 13.6 GeV or
10.7 GeV, respectively.

The Kelso/Hooper/Buckley \cite{key-1} estimate of dark matter elastic
scattering cross section with nucleons (10$^{-41}$to 10$^{-40}$
cm$^{2}$) estimates the cross section for scattering of rigid spheres
of dark matter on quark partons in nucleons boumd into atomic nuclei,
as the spheres of dark matter plow through ordinary matter in detectors.
Detectable nuclear recoils only occur in low probability events when
an incoming dark matter particle collides almost head-on with a quark
parton carrying most of the center-of-mass momentum of the struck
nucleus.
\end{onehalfspace}

\begin{onehalfspace}

\subsection*{Dark matter formation}
\end{onehalfspace}

\begin{onehalfspace}
If $\frac{\sigma}{M}$ = 0.02 cm$^{2}$/g, dark matter particle have
mass = 14.9$m_{p}$ = 14.0 GeV, radius = 1.00 fm and density $6\times10^{15}$
g/cm$^{3}$. As the universe expanded after inflation, the matter
density of the universe steadily dropped. When the matter density
in the early universe fell to $6\times10^{15}$ g/cm$^{3}$, most
matter in the universe coalesced into dark matter - small closed spherically
symmetric systems with zero orbital angular momentum. This resulted
in a universe packed with small invisible and impenetrable systems
interacting only gravitationally. Assuming uniform matter density
in the universe and instantaneous coalescence with maximum packing
fraction, 74\% of the matter in the universe is small closed impenetrable
systems constituting the bulk of dark matter. With non-uniform density,
coalescence would occur first in lower density volumes, and expansion
of the universe might allow slightly more than 74\% of matter to coalesce
into small closed systems.

The matter fraction of the energy density in the universe today is
about 0.3. If the hadronic matter fraction is about 0.05, dark matter
is about (0.25/0.3) = 83\% of all matter. If 74\% of all matter is
small closed systems, those small bound systems account for $0.74\times0.3=0.22$
of the energy density in the universe today, or about $\left(0.22/0.25\right)=88\%$
of the dark matter. Using today's scale factor of the universe $R_{0}\approx10^{28}$
cm and today's matter density $2.4\times10^{-30}$ g/cm$^{3}$, the
matter density $6\times10^{15}$ g/cm$^{3}$ required for coalescence
into small closed systems with mass 14.0 GeV occurred when the scale
factor of the universe was $R_{c}\approx7\times10^{12}$ cm.
\end{onehalfspace}

\begin{onehalfspace}

\subsection*{Minimum black hole mass}
\end{onehalfspace}

\begin{onehalfspace}
The impenetrable spheres of dark matter are the ultimate defense against
gravitational collapse. The radius of a close-packed sphere of $n$
dark matter particles is $R_{n}=\sqrt[3]{n}\frac{\hbar M}{\pi cm_{p}^{2}}=\sqrt[3]{n}$
fm. The Schwarzschild radius of that sphere, $R_{S}=\frac{29.9Gnm_{p}}{c^{2}}=3.7n\times10^{-51}$cm,
is smaller than the physical radius of the sphere until $\sqrt[3]{n}(1\times10^{-13}$cm)
$=3.7n\times10^{-51}$cm, or $n=1.4\times10^{56}$. This indicates
a minimum mass for accretionary black holes of $2.1\times10^{57}m_{p}=3.5\times10^{33}$
g, or about 1.5 times the solar mass, and a minimum Schwarzschild
radius of 5.2 km. The limit prohibits black holes with mass around
$10^{20}$ g proposed as \textquotedblleft{}anti-matter factories\textquotedblright{}
by Bambi et al \cite{key-11}.
\end{onehalfspace}

\begin{onehalfspace}

\subsection*{Accelerator production}
\end{onehalfspace}

\begin{onehalfspace}
The impenetrable spheres of dark matter will not be created in particle
accelerators because of Lorentz contraction of the accelerated particles.
In the colliding particle center of mass system, the energy of colliding
particles is dumped into a thin Lorentz-contracted disk. This does
not create the uniform matter distribution in a sphere of radius 1
fm necessary to reproduce early universe conditions resulting in coalescence
of impenetrable spheres of dark matter.
\end{onehalfspace}

\end{document}